# The general electronic principle driving size-dependent surface chemical activities of nanomaterials

Guolei Xiang[1*], Yang-gang Wang[2]

1. State Key Laboratory of Chemical Resource Engineering, College of Chemistry, Beijing University of Chemical Technology, Beijing 100029, P.R. China.
2. Department of Chemistry, Southern University of Science and Technology, Shenzhen, Guangdong, 518000, P.R. China.

Correspondence: xianggl@mail.buct.edu.cn

**Abstract:**

Size can widely affect the surface chemical activities (SCAs) of nanomaterials in chemisorption, catalysis, surface effects, etc., but the underlying electronic nature has long remained mysterious. We report a general electronic principle that drives the origin of size-dependent SCAs by combining experimental probing and theoretical modeling. Using the chemisorption of $H_2O_2$ on $TiO_2$ as a model reaction, we experimentally reveal that the central electronic process of surface chemical interactions lies in the competitive redistribution of surface atomic orbitals from energy band states into surface coordination bonds. By defining orbital potential, a site-dependent intrinsic electronic property that determines surface activities, we further establish a mathematical model to uncover the physical nature of how structural factors correlate to SCAs, particularly the roles of size. We discover that the electronic nature of size effect lies in its inverse correlation to orbital potential and amplification effect on other structural factors like defects and coordination numbers.

Nanomaterials smaller than 5 nm can widely display size-dependent surface chemical activities (SCAs) in chemisorption, catalysis and surface ligation(*1-7*). Such nano effects, represented by gold catalysis, have highly accelerated the rise and development of nanoscience and nanotechnology(*8-11*). However, the underlying electronic-level physical and chemical nature of why and how size reduction can increase and even dominate SCAs has long remained unclear (*12-16*) Some ascribed the issue to the geometrical-level factors of reduced size, increased surface area or low-coordinated surface sites(*17, 18*), while some to the atomic-level factors of defect, strain, etc(*12*). Without clarifying how the intrinsic electronic properties are modified, however, these factors cannot reveal the primordial mechanisms of these phenomena. The electronic insights are limited in three aspects. (i) Surface chemical interactions occur through forming surface coordination bonds, which can just locally perturb the bonding structures of the active site atoms. Thus, experimentally, the unperturbed electronic signals of the dominant inner atoms inevitably interfere the effective detections of electronic-level adsorbate-surface interactions.(*19*) (ii) Computationally, most nanostructures are too large systems to explore size effects on SCAs through computer simulations based on quantum mechanics(*15, 16, 20, 21*). (iii) Theoretically, physical models, concepts and theories correlating size to the critical feature of electronic structures and SCA have not been systematically developed(*14, 22, 23*). Now this issue is still a multi-disciplinary fundamental challenge in physics, chemistry, catalysis, nanoscience, etc.

   Combining experimental characterization and theoretical analysis, herein we explore the general electronic principle that drives the origin of nanoscale size-dependent surface chemical activities. We first experimentally probe the core electronic feature of surface chemical interactions with near-edge X-ray absorption spectroscopy (NEXAS), which lies in the competitive redistribution of surface valence atomic orbitals (SVAOs) of the active site atoms from extended energy band states to localized surface chemical bonds. Based on this discovery, we further define a concept, orbital potential, to describe the intrinsic competitive capabilities of surface atoms and adsorbates in forming surface chemical bonds, and further mathematically model such orbital redistribution processes to reveal the general physical nature of structure effects on SCAs, in particularly the roles of size.

   We use the chemisorption of hydrogen peroxide ($H_2O_2$) with titania ($TiO_2$) as a model reaction to experimentally probe the effects of size on SCAs. Our primary material model is $TiO_2$(B) nanosheet (NS) prepared by hydrolyzing $TiCl_4$ in ethylene glycol (Fig. 1A and S1)(*24*). The thickness is ~ 0.40 nm as shown by the side views characterized with high-resolution transmission electron microscope (HRTEM, Fig. 1B). The controls are $TiO_2$(B) nanowires (NWs) with diameters larger than 50 nm (Fig. 1C and S2), 10-nm anatase $TiO_2$ nanoparticles (NPs) prepared by annealing $TiO_2$(B) NSs at 400 ºC in air (NS-400, Fig. 1D and S3) and 15-nm anatase nanoparticles (15-nm-A, Fig. S4). On mixing with 3% aqueous $H_2O_2$ solution, the nanosheets instantly change from white into yellow. The color can be kept after being washed with water and dried at 60 ºC, and remain stable for at-least 1 year at room temperature (Fig. 1E). As characterized by solid-state UV-vis diffuse reflection spectra (DRS, Fig. 1E), this change corresponds to a new absorption mode between 2.0 and 3.3 eV. In essence, this color indicates strong orbital interactions between peroxide ligands and surface Ti atoms that allow photon-excited ligand-to-metal electron transfer, which means effective occurrence of chemisorption. We further find that sub-5-nm nanoparticles of anatase, rutile and $TiO_2$-B can all react with $H_2O_2$ to become yellow (Fig.

S5), while the controls, TiO$_2$(B) NWs, NS-400 and 15-nm-A lose the activities, as shown by the unchanged colors after peroxide modification and DRS curves (Fig. 1E). Through studying the performances of anatase nanoparticles with mean diameters of 1.6, 3.0, 5.0, 10, 15 and 21 nm (Fig. S6), we find that the critical size is between 5 and 10 nm (Fig. S7). Therefore, the surface activity of TiO$_2$ to chemisorb H$_2$O$_2$ is independent upon the phases and synthetic methods, while size is the critical factor.

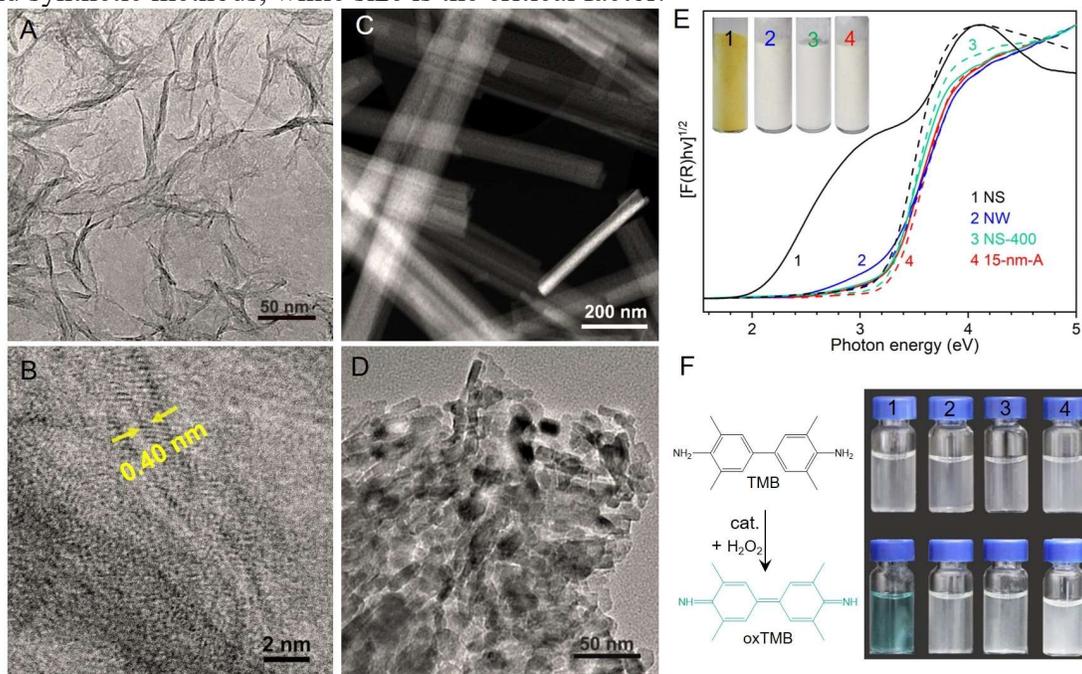

**Fig. 1. Size-dependent chemisorptive activities of peroxide on TiO$_2$ nanostructures.** (**A**) TEM image of TiO$_2$(B) nanosheets (NS). (**B**) Side view of TiO$_2$(B) nanosheets showing a thickness of ~ 0.40 nm. (**C**) TEM image of TiO$_2$(B) nanowires (NW) with diameters larger than 50 nm. (**D**) TEM image of annealed TiO$_2$(B) nanosheets at 400 ºC in air for 2 h (NS-400). (**E**) UV-vis spectra showing the absorption modes of TiO$_2$ samples before and after peroxide modification and the images of modified samples. (**F**) Images indicating the different catalytic activities of TiO$_2$ samples in the oxidation of TMB with H$_2$O$_2$.

In addition to enhanced chemisorption, size reduction can also increase the catalytic capability of TiO$_2$. We use the catalytic oxidation of 3,5,3',5'-tetramethylbenzidine (TMB) with H$_2$O$_2$, a chromogenic reaction widely used in enzyme-linked bioanalysis, to visually probe size effects on catalysis (Fig. 1F)(25). Without catalysts, H$_2$O$_2$ cannot effectively oxidize TMB, and TiO$_2$(B) NWs and anatase NPs (10 nm and 15 nm) cannot induce color change after reactions at 40 ºC for 6 h. While TiO$_2$(B) NSs can catalyze the oxidation of TMB into blue products within 10 min. This difference further shows that size reduction can lead to enhanced surface catalytic activity.

All surface chemical interactions occur based on the formation of surface coordination bonds between adsorbates and surface atoms(26, 27). We experimentally reveal the underlying core electronic process with near edge X-ray absorption fine structure (NEXAFS), a technology probing the densities of unoccupied electronic states(19). To maximize the tuning effects of surface ligands on surface electronic structures, two kinds of atomically thick nanosheets, TiO$_2$(B) and titanium oxo-peroxide (TiOP, Fig. S8), are used as the model systems. In octahedral [TiO$_6$] ligand fields, the valence atomic orbitals (AOs) of central Ti atoms (3d, 4s and 4p) overlap with the 2s and 2p AOs of the six coordinate O atoms to form molecular orbitals (MOs) (Fig. 2A). The occupied bonding MOs are composed of the AOs of O atoms, and the unoccupied antibonding MOs of the AOs of Ti atoms. These primary MOs further extend into the lattices to form sub-bands in the energy bands of TiO$_2$ (Fig. 2B)(*28*). For TiO$_2$, the most critical electronic feature regarding surface chemical interactions correlates to the *d* bands locating at the bottom of the conduction band, which include the unoccupied $\pi^*$-type $t_{2g}$ and $\sigma^*$-type $e_g$ sub-bands. Changes in the widths and intensities of the $t_{2g}$ and $e_g$ peaks of NEXAFS Ti-L$_3$ line indicate the bonding features of ligands with surface Ti atoms(*19*).

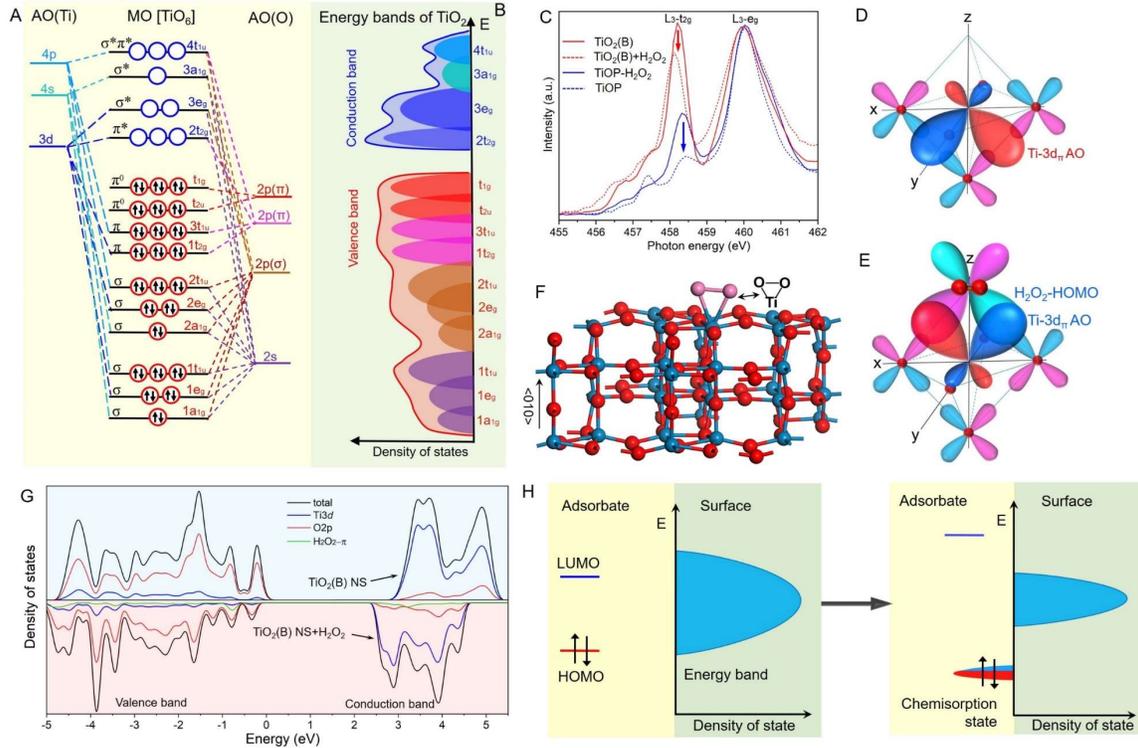

**Fig. 2. Electronic interaction through orbital redistribution between peroxide and TiO$_2$ nanosheets.** (**A**) The diagram of molecular orbitals of an octahedral [TiO$_6$] ligand field. (**B**) Fine structures of TiO$_2$ energy bands. (**C**) NEXAFS Ti-L$_3$ lines of TiO$_2$ nanosheets tuned by peroxide modification, in which the arrows point to the modified samples. (**D**) The distribution of $3d_\pi$ AOs of surface Ti atoms without surface ligands, which polarize into the lattice to enhance $t_{2g}$ energy band. (**E**) The redistribution of surface $3d_\pi$ AOs induced by peroxide ligands, which polarize into the surface chemical bonds. (**F**) The adsorption configuration of peroxide ligand on the (010) surface of one-unit-thick TiO$_2$(B) nanosheet for DFT calculations. Blue balls are Ti atoms and red ones are O. (**G**) Calculated DOSs of pure TiO$_2$(B) and peroxide-modified TiO$_2$(B) nanosheets showing ligand-induced redistribution of Ti 3d orbital and the effects on the total electronic structure of the energy bands. (**H**) The general orbital picture of chemisorption and the weakening effect on energy bands at the nanoscale.

Both the nanosheets without surface ligands display distinct $t_{2g}$ and $e_g$ peaks in the L$_3$ lines, while the $t_{2g}$ peaks are obviously depressed after peroxide modification (Fig. 2C, S9).

The changes mean that peroxide ligands can strongly interact with the $3d_\pi$ ($d_{xz}$ and $d_{yz}$) AOs of surface Ti atoms, and can modify the electronic energy bands. For $TiO_2$ nanosheets without surface ligands, the $3d_\pi$ AOs mainly extend into lattice to form Bloch states and contribute to $t_{2g}$ bands (Fig. 2D), which is indicated by the higher $t_{2g}$ peaks in Fig. 2C. While the depressed $L_3$-$t_{2g}$ peaks mean that the densities of the delocalized $t_{2g}$ states are decreased by surface peroxide ligands. The underlying electronic process lies in redistributing the $3d_\pi$ AOs from delocalized $t_{2g}$ bands into localized surface coordination bonds, which is fundamentally driven by the matched phase symmetries between the $3d_\pi$ AOs and the highest occupied molecular orbitals (HOMOs) of peroxides (Fig. 2E).

The direct experimental detections reveal that peroxides bond to surface Ti atoms through $\pi$ coordination, and act as bidentate ligands as shown in Fig. 2F. Based on this bidentate adsorption model, we further investigate the manipulation effect of peroxide ligands on the electronic structures of the one-unit-thick $TiO_2(B)$ nanosheets exposing (010) facet through theoretical calculations based on density functional theory (DFT). Figure 2G presents the total densities of states (DOS) of pure and peroxide-modified $TiO_2(B)$ nanosheets, and the projected densities of states (PDOS) of O-2p and Ti-3d AOs. After peroxide modification, the total DOS and PDOS of Ti-3d AOs are depressed at the bottom of conduction band, matching our experimental results (Fig. 2C). Therefore, both the experimental and calculation results show that adsorbates can redistribute the surface valence atomic orbitals (SVAOs) from energy band states into localized surface coordination bonds (Fig. 2H), which is the central electronic process underlying surface chemical interactions. At the nanoscale, such orbital redistributions can narrow the electronic energy bands, and further alter other physical and chemical properties.

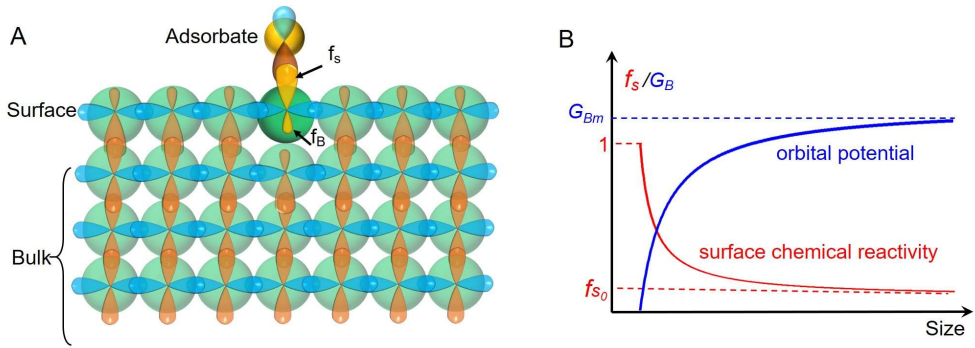

**Fig. 3. Theoretically modeling the electronic principle of size-dependent surface chemical activity.** (**A**) Scheme showing adsorbate-induced orbital redistribution in chemisorption at a specific surface site. (**B**) The dependence of orbital potential and surface chemical activity on size at the nanoscale.

Surface chemical activity positively correlates to the strength of chemisorption, which can be measured by the redistribution degrees of SVAOs. To reveal the general electronic principle underlying size-dependent surface chemical activity, we define a parameter, distribution fraction ($f$) of an atomic orbital, to describe the bonding states of a SVAOs. The physical nature of $f$ is the square of an electronic wavefunction(*19*). For a surface atom, its SVAOs can extend into Bloch states of the lattice to contribute to energy bands or localize into surface coordination bonds to form chemisorption. We use $f_B$ to denote the fraction into energy bands, and $f_S$ into a chemisorption bond (Fig. 3A). Then for fixed adsorbate and surface, $f_S$ is a descriptor measuring the strength of chemisorption.

The general structure-SCA relationship can be mathematically expressed as
$$f_S = f_S(\vec{x}) \quad (1),$$
where $\vec{x}$ represents all structural factors. Then the effect of any structural factor, $x$, on SCA can be expressed by the derivative
$$E(x) = \frac{\partial f_S}{\partial x} \quad (2).$$
Specifically, to understand the physical nature of size effect on SCA lies in revealing the function of
$$f_S = f_S(r) \quad (3).$$

According to the quantum normalization postulation of electronic wavefunctions, for a SVAO, we always have
$$f_S + f_B = 1 \quad (4).$$
This normalization equation suggests the competitive nature of $f_S$ and $f_B$ in chemisorption. For a steady state, we can have a fixed ratio of $f_S/f_B$. The value is ultimately determined by the capabilities of adsorbates and surface active sites to form chemical bonds with one SVAO. We introduce another concept, orbital potential ($G$), to describe the competitive capabilities in chemisorption. For solid surfaces, orbital potential is a site-dependent intrinsic electronic property that fundamentally determines SCA at a specific surface site. Then we have
$$\frac{f_S}{G_S} = \frac{f_B}{G_B} \quad (5),$$
where $G_B$ is the orbital potential applied to the SVAO of a specific surface site by the whole solid phase, and $G_S$ is the orbital potential of an adsorbate applied to the same SVAO. Solving eq. (4) and (5) yields
$$f_S = \frac{G_S}{G_S + G_B} \quad (6)$$
and
$$f_B = \frac{G_B}{G_S + G_B} \quad (7).$$

Equation (6) indicates that the strength of chemisorption can be enhanced by decreasing the ratio of $G_B/G_S$, that is to increase $G_S$ or to decrease $G_B$. Basically, $G_B$ and $f_S$ are functions of all structural parameters, including composition, long-range and local atomic structures, facet, particle size, adsorbate, defect, strain, coordination number, etc.

Equation (1) can be expanded as
$$f_S = f_{S_0} + \Sigma \int \frac{\partial f_S}{\partial x_i} dx_i \quad (8),$$
where $f_{S_0}$ denotes the intrinsic SCA of a specific adsorbate on the macroscale single crystal of a given material. With eq. (6), eq. (8) can be written as
$$f_S = f_{S_0} + \Sigma \int \left( \frac{\partial f_S}{\partial G_S} \cdot \frac{\partial G_S}{\partial x_i} + \frac{\partial f_S}{\partial G_B} \cdot \frac{\partial G_B}{\partial x_i} \right) dx_i \quad (9).$$
Equation (9) can be further rearranged as
$$f_S = f_{S_0} + \Sigma \int \frac{G_S G_B}{(G_S + G_B)^2} \left( \frac{\dot{G}_S}{G_S} - \frac{\dot{G}_B}{G_B} \right) dx_i = f_{S_0} + \Sigma \int f_S f_B \left( \frac{\dot{G}_S}{G_S} - \frac{\dot{G}_B}{G_B} \right) dx_i \quad (10).$$
Then the mathematic expression of structure effects on SCA is
$$E(x) = f_S f_B \left( \frac{\dot{G}_S}{G_S} - \frac{\dot{G}_B}{G_B} \right) \quad (11).$$

Equation (11) transforms the effects of structural parameters on SCA to orbital potentials. We define

$$X = \frac{\dot{G}_S}{G_S} - \frac{\dot{G}_B}{G_B} \quad (12).$$

As $0 \leq f_S, f_B \leq 1$, the plus or minus of eq. (11) depends on the sign of $X$. Thus $X$ is an electronic-level indicator determining how structural factors influence SCA. For fixed adsorbates, $\dot{G}_S = 0$, then

$$X = -\frac{\dot{G}_B}{G_B} \quad (13).$$

In general, negative $\dot{G}_B$ yields positive $X$, then increasing $x$ can enhance SCA, like defects; while, when $\dot{G}_B > 0$, increasing $x$ will weaken SCA, like size and coordination number.

For fixed adsorbates, $f_S$ can be further expressed as

$$f_S = \frac{f_{S_0}}{f_{S_0} + \gamma f_{B_0}} \quad (14) \text{ and}$$

$$\gamma = G_B/G_{Bm} \quad (15),$$

where $G_{Bm}$ is the maximum of $G_B$ when $r \to \infty$. The equations mean that the key point to reveal structure-SCA relationships lies in revealing how they correlate to $G_B$. For size effects, it is to reveal the function of $G_B = G_B(r)$. As the bonding strength of SVAOs to the neighboring atoms mainly contributes to the cohesion energy of surface atoms, $G_B(r)$ should be intrinsically consistent with the expression of cohesive energies (Fig. 3B) (*29, 30*), and can be mathematically expressed as

$$G_B = G_{Bm}(1 - \frac{kd_0}{r}) \quad (16),$$

where $d_0$ is the size of an atom and $k$ is a factor relating to material structures. Then we can have

$$f_S = \frac{f_{S_0}}{1 - \frac{kf_{B_0}d_0}{r}} \quad (17)$$

$$\text{and } X = -\frac{\dot{G}_B}{G_B} = -\frac{\dot{G}_B}{G_{Bm}(1 - \frac{kd_0}{r})} \quad (18).$$

As shown in Fig. 3B, eq. (17) indicates that SCA is inversely proportional to particle size, that is decreasing particle sizes can enhance surface chemical activities. This evolution trend is consistent with the empirical understanding of the dependence of SCA on surface-to-volume ratio (*S/V*) or *1/r*. When particle sizes can effectively dominate orbital potentials below certain critical values, eq. (18) shows that the effects of other structural factors, such as defect and dopant, on SCAs are coupled to size, and their roles can be increasingly amplified by decreasing particle size. This amplification effect resulting from size reduction is a nano effect on nanomaterial surface science, which is different from macroscale materials. The inverse proportion and amplification effect reveal the general electronic nature of size-dependent surface chemical activity.

The above theoretical models of eq. (17) and (18) are based on eq. (16), and are applicable to three-dimensionally extended materials, while layer-structured materials are exceptions. Taking layer-structured $Cs_2Ti_6O_{13}$ as an example, the $[Ti_6O_{13}]$ layers are isolated by intercalated $Cs^+$ cations (Fig. 4A). The AOs mainly delocalize within each layer, and the electronic interactions between the adjacent layers are weak. We prepare $Cs_2Ti_6O_{13}$ microcrystals by annealing $Cs_2CO_3$ and $TiO_2$ at 800 °C (Fig. 4B and S10) (*30*). Although the particles are larger than 500 nm, they can react with $H_2O_2$ to become yellow (Fig. 4C), and the DRS spectra also display a strong visible-light absorption mode resulting from the chemisorption of peroxide. In addition, layer-structured $K_2Ti_4O_9$ microcrystals, prepared via a solid-state reaction at 960 °C, can also react with $H_2O_2$ to become yellow (Fig. S11). These results mean that the intrinsic surface chemical activities of layer-structured materials are not size-dependent. Therefore, the general electronic principle underlying surface chemical activity lies in how the orbital potential of a SVAO correlates to its local bonding feature and long-range electronic states, which is the function of all structure factors at geometrical, atomic and electronic levels.

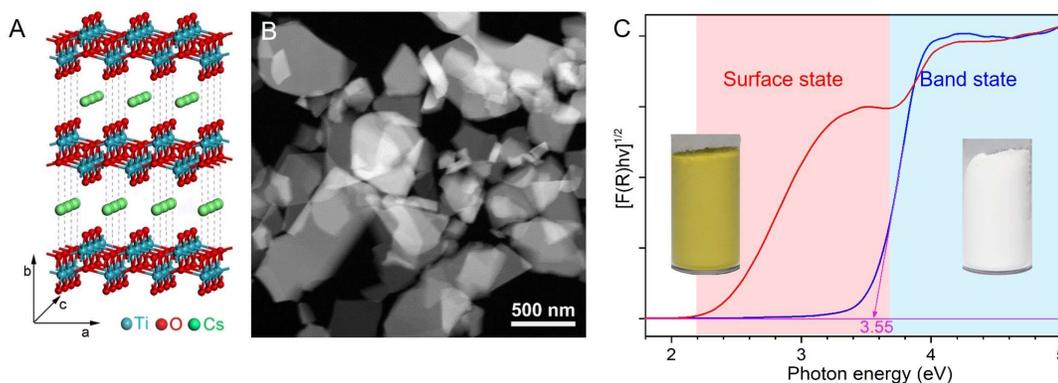

**Fig. 4. Surface reactivity of layer-structured titanate with $H_2O_2$.** (**A**) Atomic structure model of layer-structured $Cs_2Ti_6O_{13}$, in which [$TiO_6$] octahedron extends to form confined $TiO_2$ nanosheets with $Cs^+$ cations intercalated between the layers. (**B**) Scanning transmission electron microscope (STEM) image of $Cs_2Ti_6O_{13}$ crystals (>500 nm). (**C**) Optical adsorption and image of raw and peroxide-modified $Cs_2Ti_6O_{13}$ crystals.

In summary, we have revealed the electronic principle of why and how size reduction can increase and dominate the surface chemical activities of nanomaterials, through theoretically modeling the competitive orbital redistributions induced by surface coordination bonds. We transform the effects of structural factors on surface reactivity into their correlation to orbital potential, a site-dependent intrinsic electronic property determining surface reactivity. In particular, we find that the electronic nature of size-dependent surface chemical activities originates from the inverse dependence of orbital potential upon size and the amplification effects of size reduction on the roles of other structural factors. Our physical model of competitive orbital redistribution and electronic theory based on orbital potential pave a new approach to deeply explore the electronic mechanisms of structure-reactivity relationships in the fields concerning nanomaterial surface chemistry.

**ACKNOWLEDGMENTS**

This research was supported by National Natural Science Foundation of China (21801012) and China's Fundamental Research Funds for Central Universities (buctrc201812). We acknowledge the NEXAFS beamtime at BSFC granted by 2018-BEPC-PT-001799. G.X. thanks the Public Hatching Platform for Recruited Talents of Beijing University of Chemical Technology for support.

**SUPPLEMENTARY MATERIALS**

Materials and Methods

Figs. S1 to S11